\documentclass[10pt,letterpaper,oneside,notitlepage, twocolumn]{revtex4-1}
\usepackage[latin1]{inputenc}
\usepackage{mathtools}
\usepackage{amsfonts}
\usepackage{array}
\usepackage{color}
\usepackage[T1]{fontenc}
\usepackage{amssymb}
\usepackage[left=0.5in, right=0.5in, top=0.5in, bottom=0.5in]{geometry}
\usepackage[hang]{caption}
\usepackage{subcaption}
\usepackage{hyperref}
\begin{document}
\title{Using a proxy state to improve the accuracy of truncated hyperpolarizability calculations}
\date{\today}
\author{Sean Mossman}
\author{Mark G. Kuzyk}
\affiliation{Department of Physics and Astronomy, Washington State University, Pullman, Washington  99164-2814}

\begin{abstract}
    We have developed a simple algorithm for defining a single proxy state which accounts for state truncation in the sum-over-states calculations of the dispersion of the molecular  hyperpolarizabilities.  The transition strengths between the proxy state and the truncated set of states are determined using the Thomas-Reiche-Kuhn sum rules.  In addition to the sum rules, this method requires as an input the off-resonant polarizability. This proxy state method can augment experimentally determined parameters or finite-state theories to allow for a more accurate prediction of the nonlinear optical properties of molecular systems.  We benchmark this approach by comparison with exact perturbation calculations of one-dimensional power law potentials.
\end{abstract}
\maketitle

\section{Introduction}
The molecular polarizability and hyperpolarizabilities are the key microscopic quantities which characterize a given nonlinear optical material\cite{boyd09.01}. The nonlinear optical (NLO) response fundamentally mediates all photon-photon interactions and may be harnessed for a variety of photonics applications\cite{gu16.01}. While the bulk response of any given material depends on a variety of design criteria, it is limited by the quantum mechanical response of the material's microscopic constituents. Thus, there is a considerable push toward modeling, characterizing and optimizing molecular systems which will constitute the next generation of NLO devices\cite{erickson16.01, lou17.01}.

These microscopic susceptibilities are calculable from perturbation theory\cite{orr71.01} with knowledge of the many-electron transition elements $x^i_{nm}=\langle n|x^i|m\rangle$ and the energy differences $E_{nm} = E_n-E_m$, where $x^i$ is the $i^\text{th}$ Cartesian displacement operator and the state indices $n$ and $m$ represent energy eigenstates of the molecular system. The linear polarizability and hyperpolarizability can be written in terms of sum over states (SOS) expressions as
\begin{align}
    \alpha_{ij}(-\omega;\omega) = e^2\mathcal{P}_\mathcal{F}\sum_{n=1}^\infty \frac{x^i_{0n}x^j_{n0}}{E_{n0}-i\Gamma_{n0}-\hbar\omega},
    \label{eq:alphasos}
\end{align}
and
\begin{align}
    &\beta_{ijk}(-\omega_\sigma;\omega_1,\omega_2)\\
    &\hspace{0.1cm}= \frac{e^3}{2}\mathcal{P}_\mathcal{F}\sum_{n,m=1}^\infty \frac{x^i_{0n}\bar{x}^j_{nm}x^k_{m0}}{(E_{n0}-i\Gamma_{n0}-\hbar\omega_\sigma)(E_{m0}-i\Gamma_{m0}-\hbar\omega_2)},\nonumber
    \label{eq:betasos}
\end{align}
and higher order susceptibilities are more complicated but take a similar form, where $e$ is the charge of the electron and $\mathcal{P}_\mathcal{F}$ instructs us to sum over all simultaneous permutations of tensor indices with optical frequencies, where $\omega_\sigma$ is the sum frequency of the input frequencies $\omega_i$.

Models of the nonlinear polarizabilities from linear and nonlinear absorption measurements often use Eqs.~\ref{eq:alphasos} and \ref{eq:betasos} under finite state models, often with only two or three levels\cite{kuzyk90.02,cheng91.01,ensle16.01}. For example, Ensley et al. describe how the transition strength to a given state from the ground state, $x_{i0}$, can be determined from the magnitude and frequency of an absorption peak, while the damping parameter for that transition, $\Gamma_{i0}$, can be determined from the width of the absorption peak. Conversely, these expressions have been used to constrain transition properties by taking direct measurements of the nonlinear susceptibilities\cite{andre94.01, andre95.01}. Once these measurements are complete, one would like to be able to calculate the response for the same material at different frequencies and for the higher-order nonlinear optical responses, which may be difficult to measure directly. These methods have been successful for a variety of systems, but these finite-state models often fail for systems which require higher state or continuum contributions. This paper proposes a systematic method for adding a proxy state which approximately accounts for truncation while only requiring a finite number of obtainable system parameters.

\begin{figure}
    \includegraphics{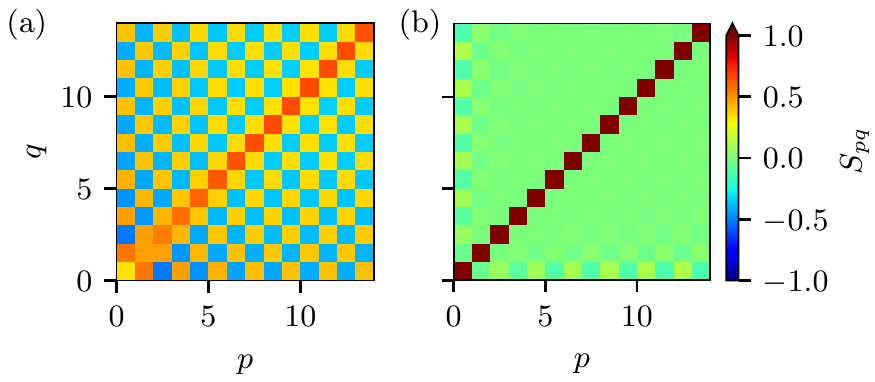}
    \caption{The truncated sum rules for the one dimensional Coulomb potential in one dimension, $V\propto -1/x$ for $x>0$, in units of $N_e\hbar^2/2m_e$ (a) using a 30 state model and (b) supplementing the 30 bound states with a single proxy state. The indexes $p$ and $q$ enumerate the states.}
    \label{fig:coulombSR}
\end{figure}

\section{Thomas-Reiche-Kuhn sum rules}
To proceed we make use of the Thomas-Reich-Kuhn (TRK) sum rules which have been used historically in discussion of absorption spectroscopy\cite{bethe77.01} and more recently as additional constraints on Eqs.~\ref{eq:alphasos}-\ref{eq:betasos} which relate the energy eigenvalues to the transition elements\cite{kuzyk00.01,kuzyk00.02, kuzyk06.01}. Along the lines of these works, the sum rules can also be used to determine difficult to measure excited state transition elements from linear absorption measurements\cite{mey12.01,perez11.02}. The sum rules are derived from the canonical commutation relation
\begin{align}
    [x,[x,H]] = \frac{\hbar^2}{2m_e},
\end{align}
where $m_e$ is the mass of the electron and $H$ can be any of a wide range of many-electron Hamiltonians\cite{kuzyk13.01}. Inserting complete sets of states yields the SOS form of the sum rules
\begin{align}
    S_{pq}=\sum_{n=0}^\infty x_{pn}x_{nq}\left(E_n-\frac{1}{2}(E_p+E_q)\right) = \frac{N_e\hbar^2}{2m_e}\delta_{pq},
    \label{eq:sumrules}
\end{align}
forming an infinite set of equations which relate the energy eigenvalues to the transition elements.

The utility of the sum rules hinges on their applicability to finite state models -- calculating an infinite sum for each of an infinite set of equations is often impractical. In 2000, a three level model was applied to both the sum rules and the SOS expressions for the nonlinear susceptibilities to obtain the fundamental limits
\begin{align}
    \alpha_\text{max} = \frac{N_e\hbar^2e^2}{m_eE_{10}^2}\text{ and }\beta_\text{max} = \sqrt[4]{3}\left(\frac{e\hbar}{\sqrt{m}}\right)^3\frac{N^{3/2}}{E_{10}^{7/2}},
    \label{eq:limits}
\end{align}
where $E_{10} = E_1-E_0$ is the energy gap between the first excited state and the ground state. For the remainder of this work the maximums stated in Eq.~\ref{eq:limits} will be used as a scale-invariant choice of units for the linear and nonlinear susceptibilities.

The sum rules as stated in Eq.~\ref{eq:sumrules} converge quickly for a wide range of systems with a well defined set of bound states, but fail for an important class of potentials in molecular design -- those with continuum states which couple to the ground state. Fig. \ref{fig:coulombSR}(a) displays each sum rule for state indices $p,q<15$ where 30 bound states are included in each sum for the 1D Coulomb potential, $V(x) = -V_0/x$ for $x>0$. The obvious deviation from the identity matrix dictated by Eq.~\ref{eq:sumrules} indicates that important states have been omitted from the sums.

To fully complete the sum rules, one must include an integral over the continuous set of unbound states with positive energy which are also admitted by potentials of the form $V\propto x^{-q}$ for $q>0$, for example. The purpose of this work is to propose an algorithm for using easily accessible information from a single polarizability measurement at zero frequency and a few bound states, along with a finite set of sum rules, to determine a single discrete proxy state which approximately accounts for the truncated bound states as well as the continuum of unbound states in the SOS expressions, allowing for more accurate dispersion calculations of the hyperpolarizabilities.

\section{The sum rule constrained proxy state}
We begin by assuming that we have a set of bound states for which we know the energy differences, $E_{nm}$, and transition elements, $x_{nm}$, where $n$ and $m$ run continuously from the ground state to a finite truncation point given by $N$. The diagonal, truncated sum rules for this set of states are given by
\begin{align}
    S^\text{bound}_{qq} = \frac{2m}{N_e\hbar^2}\sum_{n=0}^N |x_{nq}|^2 E_{nq},
    \label{eq:boundSR}
\end{align}
which must sum to values less than or equal to one if all states $q<N$ are included in the set being summed. If we then assume one additional state can be added to this set of states such that the diagonal sum rules are fully satisfied, we obtain the relation
\begin{align}
    |x_{pq}|^2E_{pq} = \frac{N_e\hbar^2}{2m_e}(1-S_{qq}^\text{bound}),
    \label{eq:proxySR}
\end{align}
where $p$ represents the proxy state and $S_{qq}^\text{bound}$ is the set of finite sums given in Eq.~\ref{eq:boundSR}. The relation Eq.~\ref{eq:proxySR} fixes the transition moments from each of the bound states to the proxy state if we have the energy of the proxy state, $E_p$, which must only be greater than all $E_q$. In fact, the off-diagonal sum rules can be greatly improved for any proxy state energy greater than $E_N$. However, to find a proxy state which is sufficiently constrained to accurately predict the nonlinear susceptibilities, we take the additional constraint to be the zero-frequency linear polarizability.

The zero-frequency linear polarizability is the first order susceptibility which relates the dipole moment resulting from an applied static field as well as the limit of the index of refraction as the frequency goes to zero. The polarizability is given by Eq.~\ref{eq:alphasos} for $\omega=0$ where we may also neglect the damping parameter, $\Gamma$. We require that the proxy state complete the polarizability sum resulting from our finite collection of bound states such that
\begin{align}
    \alpha = \alpha_\text{bound} + 2e^2\frac{|x_{0p}|^2}{E_{p0}},
    \label{eq:alphaproxy}
\end{align}
where $\alpha$ is the true zero-frequency polarizability and $\alpha_\text{bound}$ is the partial sum resulting from the finite set of bound states considered earlier.

Taking the $q=0$ case of Eq.~\ref{eq:proxySR} and inserting this into Eq.~\ref{eq:alphaproxy} allows us to solve for the energy difference between the ground state and the proxy state as a function of the difference between the truncated polarizability and the true polarizability
\begin{align}
    E_{p0} = \sqrt{\frac{N_e\hbar^2}{2m_e}\left(\frac{1-S_{00}^\text{bound}}{\alpha-\alpha_\text{bound}}\right)}.
\end{align}
With the energy of the proxy state determined by the polarizability, we can compute the transition elements $x_{pq}$ from the truncated sum rules using Eqs.~\ref{eq:proxySR}.

Fig.~\ref{fig:coulombSR}(b) shows the first 15 sum rules for the Coulomb potential in one dimension using 30 bound states and the proxy state determined by the algorithm described here. The diagonal elements are identically satisfied by design, but the off-diagonal elements are greatly improved compared with the sum rules calculated using only the bound states as shown in Fig.~\ref{fig:coulombSR}(a). This correction to the off-diagonal sum rules is insensitive to the choice of proxy-state energy, only requiring the diagonal sum rules to produce the necessary transition elements.

\section{Benchmarking the proxy state for dispersive hyperpolarizability}
To benchmark the effectiveness of the proxy state at capturing the necessary physics to correct for the state truncation in the hyperpolarizabilities calculations, we center our attention on power-law potentials in one-dimension given by
\begin{align}
    V \propto x^q \text{ for }  x>0 \text{ and } -2<q<2.
    \label{eq:powerlaws}
\end{align}
This is a reasonable space of problems as the maximum electronic hyperpolarizabilities are thought to be in quasi-one dimensional systems \cite{kuzyk13.01} and the power law potentials span a broad range of potentials including those which admit only bound state solutions as well as those which have a continuous set of scattering state solutions. We explicitly limit the space to the positive half such that the first hyperpolarizability can be nonzero.

To compare the proxy state solution with the exact solutions, we determine the polarizability and the hyperpolarizability using the exact perturbation method from Dalgarno and Lewis (DL) \cite{dalga55.01,mavro91.01}. This method requires only the ground state wavefunction to fully determine the perturbation theory by integration in position space and was recently extended to calculate the frequency dependence of nonlinear optical susceptibilities in one-dimension\cite{mossm16.01}.

First, we consider the Coulomb potential in one dimension restricted to positive space, as was used earlier as an example. This case is particularly interesting as it closely resembles the potential an electron may feel in a central potential and contains both discrete bound states as well as continuous scattering states.

\begin{figure}[t]
    \includegraphics{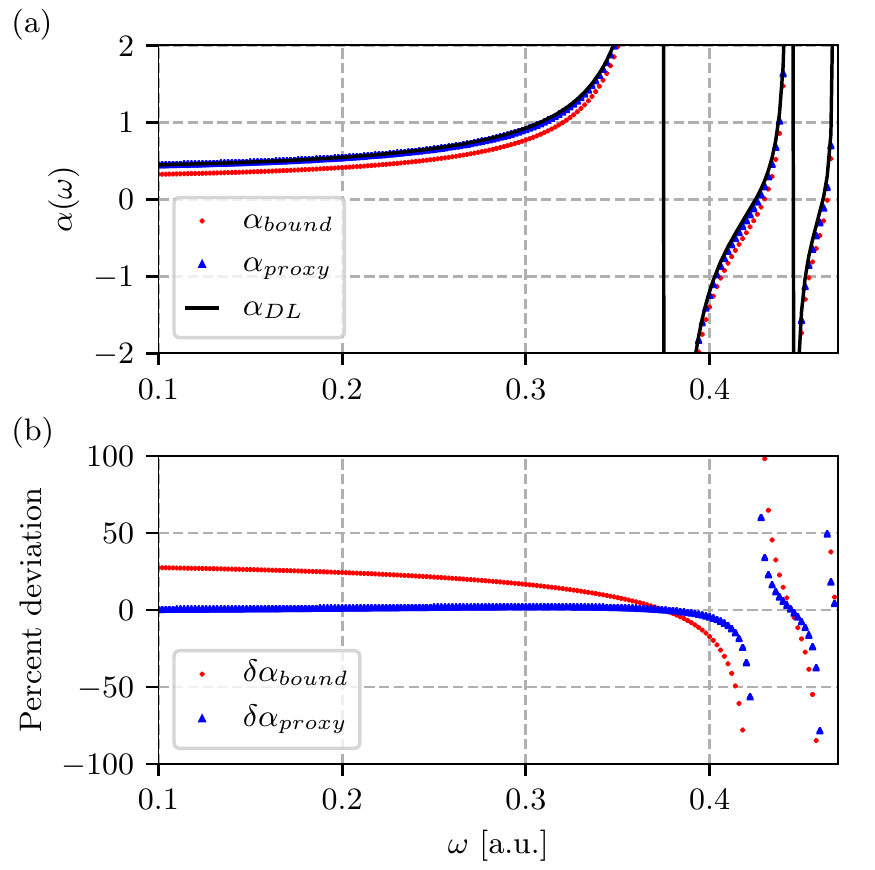}
    \caption{(a) The dispersion of the linear polarizability for a 30 state sum over states and with a single proxy state included compared with the full result calculated using Dalgarno-Lewis (DL). (b) The percent difference from the DL result for the 30 state model and for the result including the proxy state. The polarizabilities are reported in units of the off-resonant limit from Eq.~\ref{eq:limits}.}
    \label{fig:alphaproxydisp}
\end{figure}

\begin{figure}[t]
    \includegraphics{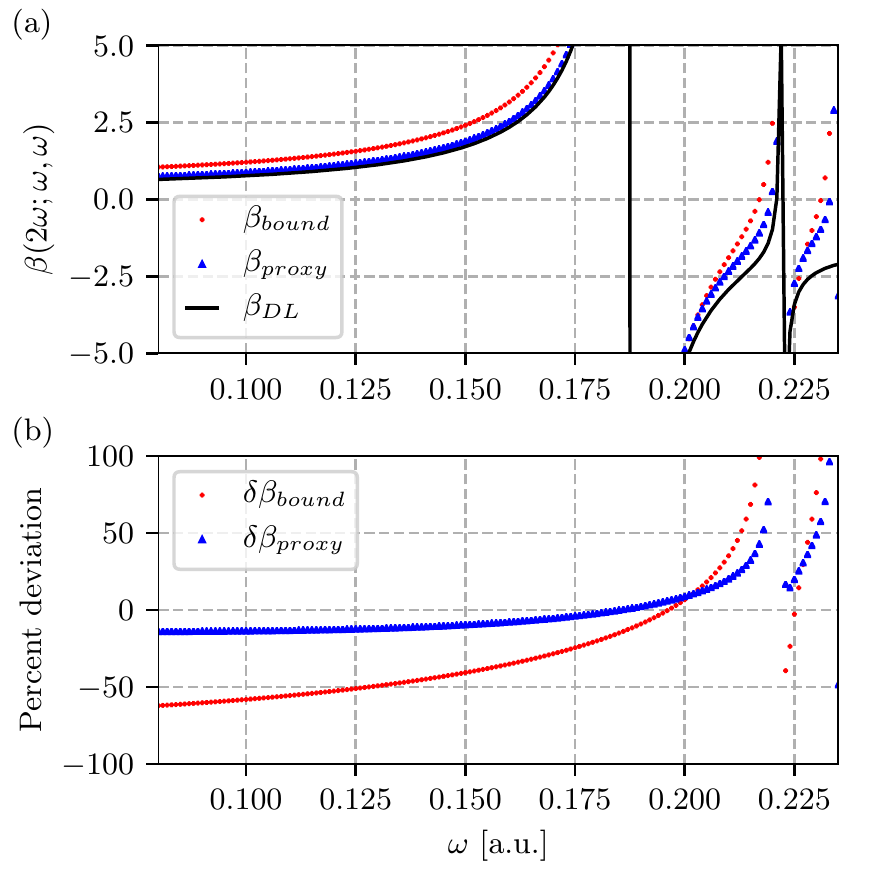}
    \caption{(a) The dispersion of the first hyperpolarizability for a 30 state sum over states and with a single proxy state included compared with the full result calculated using Dalgarno-Lewis (DL). (b) The percent difference from the DL result for the 30 state model and for the result including the proxy state. The hyperpolarizabilities are reported in units of the off-resonant limit from Eq.~\ref{eq:limits}.}
    \label{fig:betaproxydisp}
\end{figure}

Fig.~\ref{fig:alphaproxydisp}(a) and Fig.~\ref{fig:betaproxydisp}(a) show a comparison between the linear polarizability and first hyperpolarizability, respectively, obtained from the truncated set of 30 bound states, the truncated set of bound states with the proxy state, and the exact result obtained from DL. The DL solution is exact up to numerical error in the integration routine for zero damping, which limits our description to the real parts of the susceptibilities. Fig.~\ref{fig:alphaproxydisp}(b) and Fig.~\ref{fig:betaproxydisp}(b) show the percent deviation from the exact solution for both the truncated set and the truncated set with the proxy state included. For both the linear response and the nonlinear response we see significant improvement in the agreement with the exact result when including the proxy state, including near the first few resonances. The zero-frequency hyperpolarizability deviates from the true result by 68.4\% without the proxy state, and only by 13.7\% with the proxy state included. On resonance, the resulting susceptibilities are heavily dominated by the resonant state itself, and therefore, the need to correct truncation errors goes away. The percent errors spike where the susceptibilities go through zero, as one might expect.

Next, we consider the space of power law potentials described in Eq.~\ref{eq:powerlaws} where we choose the coefficient such that bound states always exist, that is to say that the potential is negative for $q<0$ and positive for $q>0$. Due to the choice of intrinsic units as described by Eq.~\ref{eq:limits}, each power law potential has a universal susceptibility regardless of the strength of the potential -- this is an equivalent statement to claiming that the intrinsic susceptibilities are scale invariant.

\begin{figure}[t]
    \includegraphics{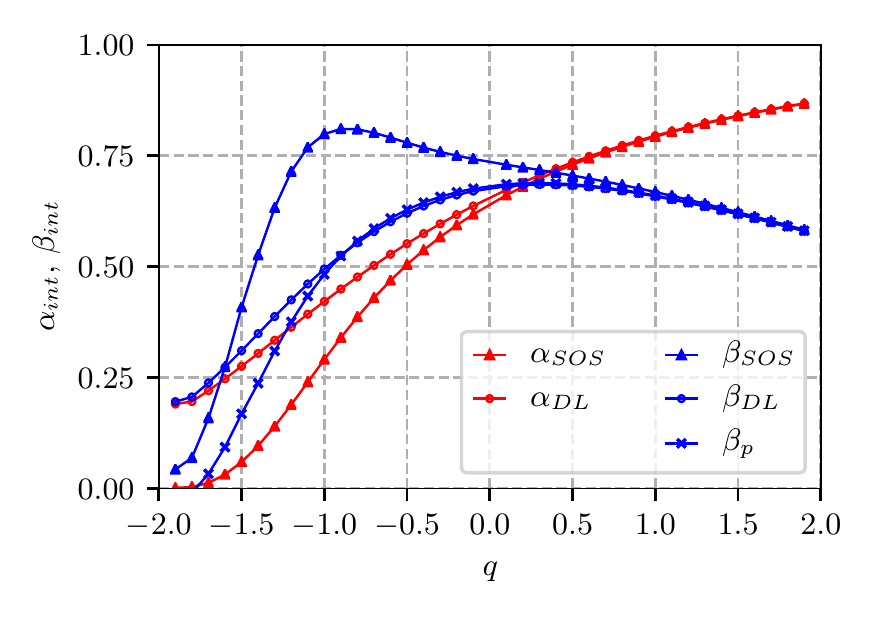}
    \caption{The off-resonant linear polarizability (red) and the first hyperpolarizability (blue) for potentials of the form $V \propto x^q$, comparing the sum-over-states (SOS) result for a 5 state model, the true result using Dalgarno-Lewis (DL), and the SOS hyperpolarizability result including the proxy state.}
    \label{fig:betaDLvSOSvProxy}
\end{figure}

To comment on all of these systems, we limit our focus to the zero-frequency susceptibilities with five bound states determined from the potential numerically using finite-differences. This must be applied quite carefully for the strongly singular potentials as the classical turning points for the solutions begin to expand exponentially as the power-law exponent approaches $q=-2$. Fig.~\ref{fig:betaDLvSOSvProxy} shows how the linear polarizability calculated with the five state model compares with the exact result. No comparison with the proxy state is shown here as the algorithm we employ would identically force agreement for the zero-frequency results discussed here. The most error due to truncation exists for the systems which admit unbound states -- the truncation error is minimal for infinitely bound systems.

Fig.~\ref{fig:betaDLvSOSvProxy} also shows the five state hyperpolarizability, the exact result from DL, and the five state model with the proxy state included. Here we continue to see little effect of truncation on the strongly bound potentials, but we see significant deviations for the $q<0$ systems. The proxy state determined from only five bound states does a remarkable job of correcting the truncation error up to potentials as singular as the Coulomb potential where the uncorrected SOS result shows qualitatively different behavior in this regime. This off-resonant result shows that the proxy state algorithm presented here is particularly effective at capturing the contributions to the nonlinear optical susceptibilities from the unbound states which couple to the ground state.

\section{Conclusion}
We have shown that with a finite number of measurable quantities, one can use the TRK sum rules to produce a proxy state which accounts for truncation of the SOS perturbation calculation for the nonlinear optical susceptibilities. This algorithm requires a finite set of well determined transition elements and energies which can be determined from linear spectroscopic measurements as well as the off-resonant linear polarizability.

We have shown that for the test case of a Coulomb potential confined to the positive side of one-dimensional space, this proxy state algorithm can significantly reduce the truncation error attributed to the unbound states of the system throughout the dispersion of the linear and first nonlinear susceptibilities. This test case shares many of the features of more physical molecular systems and is therefore evidence that this algorithm may provide significant improvements for careful calculations of the nonlinear susceptibilities. Though we have focused here on the polarizability and the first hyperpolarizability, this algorithm simply generates an additional transition element and can thus be applied to any SOS calculation of higher order susceptibilities or any other perturbative theories involving position transition elements. This method may be of particular use, with little effort, to experimental or numerical investigations of complex molecular systems where only the first few molecular orbitals may be within reach. While the benchmarks used in this work to motivate the efficacy of this approach have focused on one-dimensional problems, the TRK sum rules easily generalize to three dimensions and therefore this algorithm can be applied to three-dimensional problems with knowledge of the diagonal elements of the polarizability tensor.

Finally, we have shown the distinct correction obtained for the class of singular power law potentials when applying the proxy state correction. This shows the particularly usefulness of this algorithm when the system in question contains bound states which closely resemble and rapidly approach in energy those unbound states near zero-energy. While this likely has the largest impact on theoretical studies of the nonlinear optical susceptibilities from first principles, these concepts may also find use in practical cases where continuum states are thought to contribute to the nonlinear optical response but are difficult to calculate explicitly.

\section{Acknowledgements}
National Science Foundation (NSF) (ECCS-1128076)

We acknowledge fruitful discussion with Yan Yangqian and Doerte Blume regarding the continuum states of the one-dimensional Coulomb potential. We also acknowledge invigorating discussion with Rick Lytel.

\end{document}